\input harvmac 
\input epsf.tex

\overfullrule=0mm
\def\file#1{#1}
\newcount\figno \figno=0
\newcount\figtotno
\figtotno=0
\newdimen\captionindent
\captionindent=1cm
\def\figbox#1#2{\epsfxsize=#1\vcenter{
\epsfbox{\file{#2}}}}
\newcount\figno
\figno=0
\def\fig#1#2#3{
\par\begingroup\parindent=0pt\leftskip=1cm\rightskip=1cm\parindent=0pt
\baselineskip=11pt
\global\advance\figno by 1
\midinsert
\epsfxsize=#3
\centerline{\epsfbox{#2}}
\vskip 15pt
{\bf Fig. \the\figno:} #1\par
\endinsert\endgroup\par
}
\def\figlabel#1{\xdef#1{\the\figno}}
\def\encadremath#1{\vbox{\hrule\hbox{\vrule\kern8pt\vbox{\kern8pt
\hbox{$\displaystyle #1$}\kern8pt}
\kern8pt\vrule}\hrule}}
\def\encadre#1{\vbox{\hrule\hbox{\vrule\kern8pt\vbox{\kern8pt#1\kern8pt}
\kern8pt\vrule}\hrule}}


\def\IR{\relax{\rm I\kern-.18em R}}
\font\cmss=cmss10 \font\cmsss=cmss10 at 7pt
\def\IZ{\relax\ifmmode\mathchoice
{\hbox{\cmss Z\kern-.4em Z}}{\hbox{\cmss Z\kern-.4em Z}}
{\lower.9pt\hbox{\cmsss Z\kern-.4em Z}}
{\lower1.2pt\hbox{\cmsss Z\kern-.4em Z}}\else{\cmss Z\kern-.4em Z}\fi}
\def\buildrel#1\under#2{\mathrel{\mathop{\kern0pt #2}\limits_{#1}}}

\Title{SPhT/99-126}
{{\vbox {
\bigskip
\centerline{Meanders: Exact Asymptotics}
}}}
\bigskip
\centerline{P. Di Francesco,}
\medskip
\centerline{O. Golinelli}
\medskip
\centerline{and} 
\medskip
\centerline{E. Guitter\footnote*{e-mails:
philippe,golinelli,guitter@spht.saclay.cea.fr},}

\bigskip

\centerline{ \it Service de Physique Th\'eorique, C.E.A. Saclay,}
\centerline{ \it F-91191 Gif sur Yvette, France}

\vskip .5in

We conjecture that meanders are governed by the gravitational version
of a $c=-4$ two-dimensional conformal field theory, allowing for exact predictions
for the meander configuration exponent $\alpha=\sqrt{29}(\sqrt{29}+\sqrt{5})/ 12$,
and the semi-meander exponent ${\bar \alpha}=1+\sqrt{11}(\sqrt{29}+\sqrt{5})/24$. 
This result follows from an interpretation of meanders
as pairs of fully packed loops on a random surface, described by two $c=-2$
free fields. 
The above values agree with recent numerical estimates.
We generalize these results to a score of meandric numbers with various 
geometries and arbitrary loop fugacities. 

\noindent
\Date{10/99}

\nref\HMRT{K. Hoffman, K. Mehlhorn, P. Rosenstiehl and R. Tarjan, {\it
Sorting Jordan sequences in linear time using level-linked search
trees}, Information and Control {\bf 68} (1986) 170-184.}
\nref\ARNO{V. Arnold, {\it The branched covering of $CP_2 \to S_4$,
hyperbolicity and projective topology},
Siberian Math. Jour. {\bf 29} (1988) 717-726.}
\nref\KOSMO{K.H. Ko, L. Smolinsky, {\it A combinatorial matrix in
$3$-manifold theory}, Pacific. J. Math. {\bf 149} (1991) 319-336.}
\nref\TOU{J. Touchard, {\it Contributions \`a l'\'etude du probl\`eme
des timbres poste}, Canad. J. Math. {\bf 2} (1950) 385-398.}
\nref\LUN{W. Lunnon, {\it A map--folding problem},
Math. of Computation {\bf 22}
(1968) 193-199.}
\nref\NOUS{P. Di Francesco, O. Golinelli and E. Guitter, {\it Meanders: 
a direct enumeration approach}, Nuc. Phys. {\bf B 482} [FS] (1996) 497-535.}
\nref\Goli{O. Golinelli, {\it A Monte-Carlo study of meanders}, preprint 
cond-mat/9906329, to appear in EPJ {\bf B} (2000). }
\nref\JEN{I. Jensen, {\it Enumerations of Plane Meanders}, preprint 
cond-mat/9910313.}
\nref\LZ{S. Lando and A. Zvonkin, {\it Plane and Projective Meanders},
Theor. Comp.  Science {\bf 117} (1993) 227-241, and {\it Meanders},
Selecta Math. Sov. {\bf 11} (1992) 117-144.}
\nref\DGG{P. Di Francesco, O. Golinelli and E. Guitter, {\it Meander,
folding and arch statistics}, Mathl. Comput. Modelling {\bf 26} (1997) 97-147.}
\nref\MAK{Y. Makeenko {\it Strings, Matrix Models, and Meanders}, 
Nucl.Phys.Proc.Suppl. {\bf 49} (1996) 226-237.}
\nref\SSz{G. Semenoff and R. Szabo {\it Fermionic Matrix Models} Int.J.Mod.Phys. 
{\bf A12} (1997) 2135-2292.}
\nref\NTLA{P. Di Francesco, O. Golinelli and E. Guitter, {\it
Meanders and the Temperley-Lieb algebra}, Commun.Math.Phys. {\bf 186} (1997) 1-59.}
\nref\KPZ{V.G. Knizhnik, A.M. Polyakov and A.B. Zamolodchikov, Mod. Phys. Lett.
{\bf A3} (1988) 819; F. David, Mod. Phys. Lett. {\bf A3} (1988) 1651; J.
Distler and H. Kawai, Nucl. Phys. {\bf B321} (1989) 509.}
\nref\JACO{J. Jacobsen and J. Kondev, {\it Field theory of compact polymers
on the square lattice}, Nucl. Phys. {\bf B 532} [FS], (1998) 635-688,
{\it Transition from the compact to the dense phase of two-dimensional polymers}, 
J. Stat. Phys. {\bf 96}, (1999) 21-48.}
\nref\EKN{E. Guitter, C. Kristjansen and J. Nielsen, 
{\it Hamiltonian Cycles on Random Eulerian Triangulations},
Nucl. Phys. {\bf B546 [FS]} (1999) 731-750; 
P. Di Francesco, E. Guitter and C. Kristjansen,
{\it Fully Packed O(n=1) Model on Random Eulerian Triangulations},
Nucl. Phys. {\bf B549 [FS]} (1999) 657-667.}
\nref\DaDu{F. David and B. Duplantier, {\it Exact partition functions and correlation
functions of multiple Hamiltonian walks on the Manhattan lattice}, J. Stat. 
Phys. {\bf 51}, (1988) 327-434.}
\nref\CK{L. Chekhov and C. Kristjansen, {\it Hermitian Matrix Model with Plaquette 
Interaction}, Nucl.Phys. {\bf B479} (1996) 683-696.}
\nref\EKR{B. Eynard and C. Kristjansen, {\it More on the exact solution of the O(n)
model on a random lattice and an investigation of the case $|n|>2$}, Nucl. Phys.
{\bf B466 [FS]} (1996) 463-487.}
\nref\BAR{B. Durhuus, J. Fr\"olich and T. J\'onsson, Nucl. Phys. {\bf B240} (1984) 
453-480; F. David, Nucl. Phys. {\bf B487[FS]} (1997) 633-649.}


\newsec{ Introduction}

Meanders are a simply stated combinatorial problem consisting in counting the number 
$M_n$ of configurations of a closed self-avoiding road crossing an infinite river 
through a given number $2n$ of bridges. 
Meanders appear in several domains of science including computer science \HMRT, 
mathematics in connection with both Hilbert's 16th problem and the enumeration 
of ovals of planar algebraic curves \ARNO\ and the classification of 
3-manifolds \KOSMO.
Meanders also appear in physics as a particular example of critical phenomena:
indeed, meanders also count a particular class of Self-Avoiding Walks describing
the compact foldings of a linear chain.

Among the various techniques used to attack the problem
we can mention direct enumerations \TOU\ \LUN\ \NOUS\ \Goli, whose most recent 
one \JEN\ enumerates up to $2n=48$ bridges with a new transfer matrix method.
Other approaches use random matrices \LZ\ \DGG\ \MAK\ \SSz, or 
algebraic techniques based 
on the Temperley-Lieb algebra \NTLA. Several exact results have been obtained
for meander-related issues, such as exact sum rules for meandric numbers \DGG, 
and the calculation of a meander-related determinant \NTLA\ \KOSMO, but
despite many efforts, no explicit formula for $M_n$ has been found so far for 
arbitrary $n$. 

As a critical phenonemon, meanders are characterized by critical exponents
describing the asymptotic behavior of $M_n$ for large $n$. We expect 
a behavior:
\eqn\asybeh{M_n {\buildrel {n\to \infty} \under \sim} {R^{2n}\over n^\alpha}\ ,}
where ${\rm Log}\, R$ is the entropy per bridge and $\alpha$ the configuration
exponent. 
The best estimates extracted from extrapolation of finite $n$ exact results
read $R^2=12.262874(15)$ and $\alpha=3.4206(4)$ \JEN. 

In this paper, we present explicit formulas for the asymptotics of meanders
based on a conjecture stating that meanders are governed by a
two-dimensional conformal field theory with central charge $c=-4$
coupled to gravity. In particular, we obtain
\eqn\resalpha{\alpha=\sqrt{29}\, {\sqrt{29}+\sqrt{5}\over 12}= 3.42013288...\ ,}
in agreement with \JEN\  and 
\eqn\resalphabar{{\bar \alpha}=1+\sqrt{11}\, {\sqrt{29}+\sqrt{5}\over 24}=
2.05319873...\ ,}
where $\bar \alpha$ is the configuration exponent describing the asymptotics
for the semi-meander numbers $\bar M_n$ counting configurations of a closed 
self-avoiding road crossing a semi-infinite river (i.e a river with a source
around which the road may wind) through $n$ bridges. Again this value
is in agreement with the best estimate ${\bar \alpha}= 2.056(10)$
found in \Goli. 

Our conjecture is based on an interpretation of the meander problem
as a pair of two fully packed loop models on a random surface, whose
counterpart on a flat surface is a two-dimensional Coulomb gas whose
critical behaviour is described by two decoupled $c=-2$ free fields.  

The paper is organized as follows:
In Section 2, we recall the $O(n_1,n_2)$ matrix model describing 
the generating function
for meanders and semi-meanders with possibly several connected components 
of road and river, with a fugacity $n_1$ per river and $n_2$ per road.  
Section 3 identifies the matrix model as a fully packed loop problem 
on a random surface and discusses its flat counterpart, thus obtaining 
the central charge for arbitrary fugacities. In Section 4, we
extract the critical exponents $\alpha$ and $\bar \alpha$ for meanders and
semi-meanders ($n_1,n_2\to 0$) thanks to the KPZ formula \KPZ\ relating flat to 
random geometry. Section 5 presents several extensions of the configuration exponents
corresponding to more involved river geometries. We conclude with more
prospective results for arbitrary fugacities.

\newsec{Meanders and the O$(n_1,n_2)$ matrix model}

As shown in \DGG, the meander problem can be formulated as a
Hermitian matrix model, hereafter referred to as the $O(n_1,n_2)$ model,
with $n_1$ black matrices $B_1,..,B_{n_1}$
and $n_2$ white ones $W_1,...,W_{n_2}$, all of size $N\times N$, with partition
function
\eqn\bw{\eqalign{
Z_{n_1,n_2}(N;x)~&=~ \int \prod_{i=1}^ {n_1}dB_i \prod_{j=1}^{n_2} dW_j
e^{-N{\rm Tr}\, V(\{B_i\} ,\{ W_j\})}\cr
V(\{B_i\} ,\{ W_j\})~&=~ {1 \over 2}\big(\sum_{i=1}^ {n_1}B_i^2 
+\sum_{j=1}^ {n_2}W_j^2
-x \sum_{i=1}^ {n_1}\sum_{j=1}^ {n_2}B_i W_j B_i W_j\big)\ . \cr}}
Expanding the planar free energy, we get
\eqn\frenerg{\eqalign{F_{n_1,n_2}(x)&=
\lim_{N\to \infty} {1\over N^2} {\rm Log}\, Z_{n_1,n_2}(N;x)\cr
&=\sum_{n=1}^\infty {x^{n}} n_1n_2 f_{n}(n_1,n_2) \ , \cr}}
where $n_1n_2f_n(n_1,n_2)$ counts the total number of planar (genus $0$) connected 
self-avoiding but mutually intersecting black and white loop configurations 
with $n$ intersections, weighted by their inverse symmetry factor
and by a factor $n_1$ per black loop and $n_2$ per
white one. Due to planarity, $f_n=0$ for odd $n$. 

\fig{Typical planar (i.e. drawn on a sphere)
diagrams contributing to (a) $f_{8}(0,0)$
and (c) ${\bar f}_7(0,0)$, and one of their respective (semi-)meander pictures (b)
and (d).}{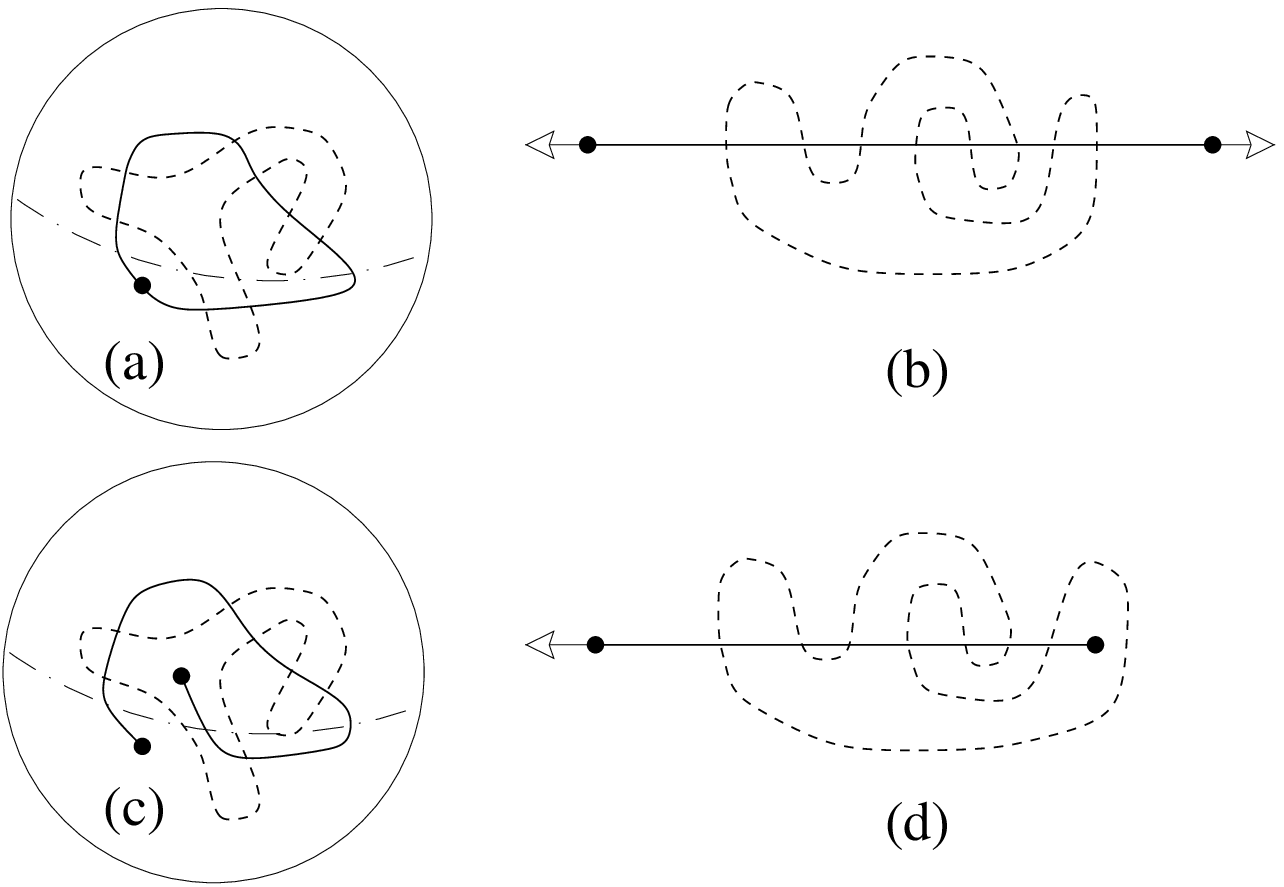}{9.cm}
\figlabel\sphere

The meander numbers $M_n$
read
\eqn\meoff{ M_n=4n f_{2n}(0,0) \ ,}
where the factor $4n$ accounts for the $2n$ positions between bridges
on the black loop where to open it and $2$ for the east-west orientation of the
meander (see Fig.\sphere\ (a) and (b)). 
In this language, rivers correspond to black loops, while roads correspond to white
ones, and the limit $n_1,n_2\to 0$ simply eliminates the configurations
with more than one loop of each color. 

A semi-meander is nothing but a configuration of a black open segment and
a white closed loop, in which one of the extremities of the black segment
is sent to infinity (this is always possible in the planar case, where configurations
are drawn on a sphere, as illustrated in Fig.\sphere\ (c) and (d)). 
In the matrix language, it corresponds to a 
large $N$ correlation function
of the operator 
\eqn\opera{\phi_1=\lim_{N\to \infty} {1\over N}\sum_{i=1}^{n_1}{\rm Tr}(B_i)}
that creates a black endpoint, namely
\eqn\semimean{
\langle \phi_1 \phi_1 \rangle =\sum_{n=1}^\infty n_1 n_2{\bar f}_n(n_1,n_2) x^n\ ,}
where $n_1 n_2{\bar f}_n(n_1,n_2)$ counts the number of configurations of rivers
made of one segment and a number of loops, intersecting closed roads.
Again the semi-meander numbers simply read
\eqn\semif{ {\bar M}_n= 2{\bar f}_n(0,0) \ ,}
as we have picked one endpoint of the segment to send it to the infinity
on the left (see Fig.\sphere\ (d)).

\newsec{Meanders as a height model: fully packed loop model}

The $O(n_1,n_2)$ model above is a particular version of a fully-packed 
loop (FPL) model on a random surface.  The random surface is dual to the 
graphs occurring in the Feynman expansion of the free energy \frenerg.  
By full packing, we mean that the loops visit
all the vertices of the graph. Moreover, each edge is visited by either a black
or a white loop.

On the regular square lattice, these two properties are characteristic of the
$FPL^2(n_1,n_2)$ loop model of \JACO.  
A configuration of this model is characterized by a set of fully packed
black loops visiting all the vertices and half of the edges, the other
half of the edges forming fully packed white loops. Each black (resp. white) loop
receives a weight $n_1$ (resp. $n_2$). 
For $n_1=n_2=2$, the loop fugacities are realized by assigning independent
orientations to all the loops.
An oriented black and white fully packed loop configuration may be equivalently
translated into a three-dimensional height configuration on the faces of
the lattice as follows.
We first bicolor the
vertices of the lattice, by letting vertices marked with
$\bullet$ and with $\circ$ alternate around each face. Next we
define a vector variable on each edge according to the rule: 
\eqn\edgovar{\eqalign{ A &\leftrightarrow \figbox{1.5cm}{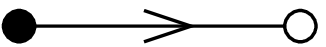} \cr
-B &\leftrightarrow \figbox{1.5cm}{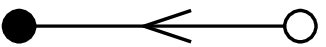} \cr
C &\leftrightarrow \figbox{1.5cm}{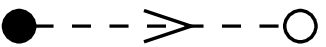} \cr
-D &\leftrightarrow \figbox{1.5cm}{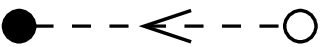} \cr}}
where $A,B,C,D$ are four fixed vectors with vanishing sum (hence 
generically in $\IR^3$).
To determine the height $h$ on each face, we use the Amp\`ere rule, that $h$
increases (resp. decreases) by the edge vector crossed, if it points to
the left (resp. right). 
This is well defined thanks to the relation $A+B+C+D=0$.
Note in this formulation that exchanging 
$A\leftrightarrow -B$ (resp. $C\leftrightarrow -D$) along a loop
amounts to reversing the orientation of the corresponding black (resp. white) 
loop.
This defines the $FPL^2(n_1=2,n_2=2)$ model, which is critical. 
In terms of the height variable, this model
is described
in the continuum limit by three free fields (one for each component of the
height vector), hence a conformal theory with central charge $c=3$. 
This model can be
modified by introducing local Boltzmann weights that assign a 
weight $n_1$ resp. $n_2$ per loop of either kind: this is the
$FPL^2(n_1,n_2)$ model.
Remarkably, the $FPL^2(n_1,n_2)$ model remains critical for $n_1,n_2\leq 2$. 
It is still described by a $3$-dimensional
Coulomb gas, but with two additional electric charges at infinity, resulting
in a central charge \JACO:
\eqn\cenchar{ c_{FPL}(n_1,n_2)=3-6({e_1^2\over 1-e_1}+{e_2^2\over 1-e_2})\ , \qquad
n_i=2\cos(\pi e_i), \ i=1,2 \ .}

We may now define this model on a random surface, by representing
its configurations as graphs made of white and black edges (for the two types
of loops), and vertices of the form
\eqn\vertifo{\eqalign{ a&=\figbox{1.5cm}{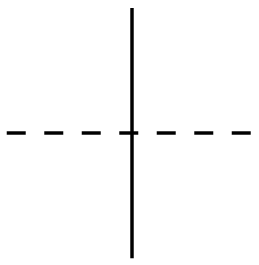} \cr 
b&=\figbox{1.5cm}{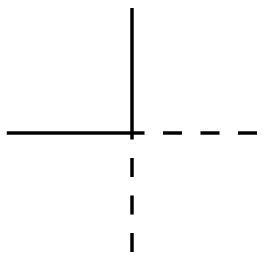} \cr}}
and weighting each black, resp. white, loop with a factor $n_1$, resp. $n_2$.
Moreover, as we have seen above, the vertices of the graph must be bicolored
with alternating marks $\bullet$ and $\circ$,
and the white and black loops
oriented, in order to define a unique three-dimensional
edge configuration, using \edgovar. 
On planar graphs
the bicolorability of the vertices ensures that the tessellation dual to the graph 
is Eulerian.

If we relax this constraint of bicolorability,
it is no longer possible
to define three-dimensional edge variables, but it is actually easy to
see that the black and white loop configurations are now faithfully reproduced
by considering only two edge variables, say $A$ for black edges and $C$ for white
ones, which in turn amounts to setting $A+B=C+D=0$. With the same Amp\`ere rule
across oriented black or white edges, we see that the height $h$ 
becomes two-dimensional, as it takes only values of the form $h_0+m A+p C$,
$m,p$ two integers.
The net effect has therefore been, by lack of bicolorability of the graphs,
to reduce the height variable to a two-dimensional
space, resulting in 
\eqn\charcen{\eqalign{
c(n_1,n_2)&= 2-6{e_1^2\over 1-e_1}-6{e_2^2\over 1-e_2}\cr &= c(n_1)+c(n_2) \cr
c(n)&= 1-6{e^2\over 1-e} \ ,\qquad
n=2\cos(\pi e) \ .\cr}}
This shift by $-1$ in the central charge when going from Eulerian to unconstrained
tessellations has already been observed in \EKN. 
Note also that the central charge $c(n_1,n_2)=c(n_1)+c(n_2)$
is that of two decoupled free fields. In flat space,
such an effective decoupling of the two {\it a priori}
coupled free fields describing the FPL model has already been observed in \JACO. 

\fig{Height configurations around a vertex of the $FPL^2(n_1,n_2)$ model
on a non-bicolorable graph.
Black edges correspond to the values $A=-B$, while white edges correspond
to $C=-D$.
In the $b$ vertex, the height on the two opposite SW and NE
faces is the same, hence the vertex may be undone as shown.}{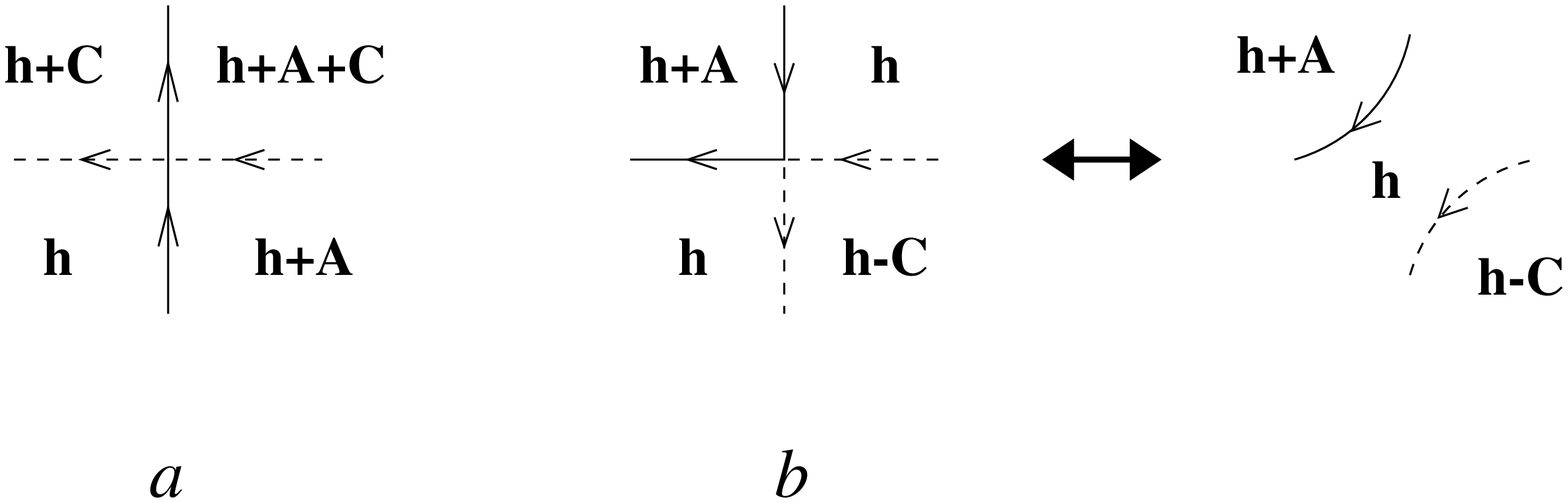}{10.cm}
\figlabel\troisd

The $O(n_1,n_2)$ model is a special version of this in which the type $b$ vertex of
\vertifo\ and Fig.\troisd\ is forbidden. As shown in Fig.\troisd, this vertex 
is expected to be irrelevant anyway,    
as the height $h$ takes the {\it same} value in the SW and NE faces, so that
the vertex can be ``undone" to let these two faces communicate without altering
the height configuration. 

\fig{Height configurations around a vertex of the $O(n_1,n_2)$ model.
Black edges correspond to the values $A$ or $-B$, while white edges correspond
to $C$ or $-D$.
The height difference between two opposite faces may only
take the values $A+C$, $A+D$, $B+C=-(A+D)$ and $B+D=-(A+C)$.}{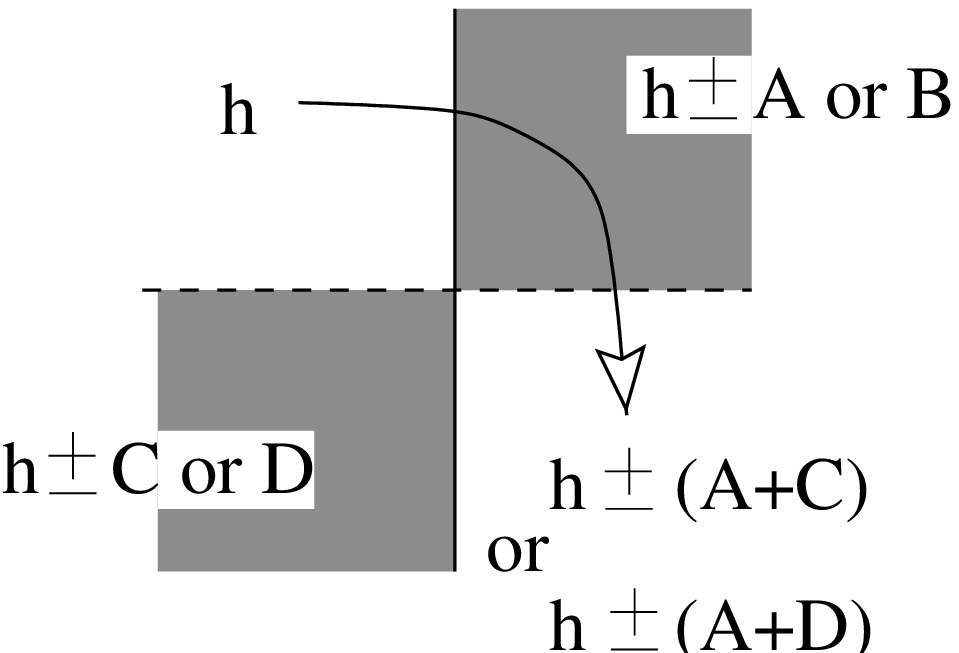}{4.cm}
\figlabel\deuxd

It is interesting to note that for this particular case (without $b$ vertex) the
bidimensionality of the height variable can be recovered in a slightly different way.
Indeed, in the absence of the type $b$ vertex, the graphs are automatically bicolorable,
hence allowing a priori for the construction of a three-dimensional height.
But
it turns out that
the heights on two opposite faces around a vertex of type $a$
may only differ by the quantities $\pm (A+C)$, $\pm (A+D)$, as illustrated in
Fig.\deuxd, whereas the differences $\pm (A+B)=\mp (C+D)$ are forbidden.
This means that the graph, whose faces can be naturally bicolored
(say black and white), must have
all the heights on white faces in the same plane generated by $A+C$, and $A+D$,
and all the heights on the black faces on a parallel plane, distant by $C$ or $D$.
For the sake of simplicity, we may take $B=-A$ and $D=-C$
as above, without altering the model since the differences $\pm (A+B)$ and $\pm (C+D)$
never appear here, and the heights all lie in the same plane generated by $A$ and $C$.

In conclusion, we are led to the natural conjecture that 
the $O(n_1,n_2)$ model is the random surface version of a critical fully packed
loop model described on the square lattice by two free fields, and with
central charge \charcen\ above.
In the particular case of meanders when $n_1=n_2=0$, i.e. $e_1=e_2={1\over 2}$, we find
$c=-2-2=-4$ as announced.

\newsec{Exact exponents from Quantum Gravity at $n_1=n_2=0$}

The above identification of the meander problem as a $c=-4$ field theory on a
random sphere can be confirmed by computing various
exact critical exponents of the $c=-4$ conformal theory coupled to 
two-dimensional gravity, as expressed through the
celebrated KPZ formula \KPZ, relating the anomalous dimensions of operators in
the lattice and random surface versions of the theory.

Defining a conformal theory on a random surface, one is led to introduce
a new parameter, the cosmological constant, coupled to the area
of the surfaces. In our matrix language, its role is played by the parameter $x$
in \bw, through the weight $x^{n}$, where $n$ is the total number of vertices
of the random graph, as well as the total area of its dual, made of squares
of unit area.
Criticality is reached when $x\to x_c$, such that the free energy's behavior 
becomes singular, with a power law
\eqn\gammma{ F(x)\sim (x_c-x)^{2-\gamma_{str}} \ ,}
where $\gamma_{str}$ stands for the string susceptibility exponent, related to 
the central charge $c$ through \KPZ\
\eqn\kpz{ \gamma_{str}(c)= {c-1-\sqrt{(25-c)(1-c)} \over 12}\ , }
valid for all $c\leq 1$.
This is immediately translated into the asymptotic behavior of the
coefficients
$F_n$ in the expansion $F(x)=\sum f_n x^n$ as
\eqn\asympto{ f_n \sim {(x_c)^{-n} \over n^{3-\gamma_{str} } }\ . }

In the case of meanders, we have $c=-4$, and 
\eqn\meander{ M_n \sim {R^{2n} \over n^{\alpha}} \ ,}
where $R_c=1/x_c$, and
\eqn\expomean{ \alpha=2-\gamma_{str}(c=-4)= {29+\sqrt{145}\over 12}=
3.42013288... }
This value is in agreement with the recent improved numerical estimate
$\alpha=3.4206(4)$ \JEN.

As mentioned above, the semi-meander numbers involve the computation of 
a correlation function of operators inserting black endpoints. The operators
of a conformal theory are known to be dressed when the theory is coupled to
gravity, and their correlations have the following behavior when the cosmological
constant $x$
approaches its critical value $x_c$: 
\eqn\correla{ \langle \phi_{m_1} \phi_{m_2} ... \phi_{m_k}\rangle \sim
(x_c-x)^{\sum_{i=1}^k \Delta_{m_i} -\gamma_{str} -(k-2)} \ , }
where $\Delta_m$ is the anomalous dimension of the dressed operator $\phi_m$,
and $\gamma_{str}$ is as in \kpz. 
Furthermore, the dressed dimension $\Delta$ is related to the conformal dimension
$h$ of the (undressed) operator of the conformal theory through the relation \KPZ\
\eqn\twokpz{ \Delta={\sqrt{1-c+24 h} -\sqrt{1-c} \over \sqrt{25-c}-\sqrt{1-c}}\ . }
Let us now return to the case of semi-meander numbers,
given by \semimean\-\semif. The operator $\phi_1$ creating black endpoints
actually pertains to the gravitational version of the $c=-2$
free field theory describing the black loops. At this stage, we can keep an
arbitrary weight $n_1$ for the black loops, i.e. consider the $O(n_1)$ theory
with $c(n_1)=1-6e_1^2/(1-e_1)$, $n_1=2\cos(\pi e_1)$. In the corresponding
Coulomb gas language, the correlator $\langle \phi_1\phi_1\rangle$ corresponds to a 
correlation $\langle \psi_1(z_1) \psi_{-1}(z_2)\rangle$ of a 
conformal operator that creates 
an oriented dislocation line in the height picture between
two magnetic monopoles at $z_1$ and $z_2$. 
The corresponding magnetic charges $\pm m$ must be corrected by an electric 
charge $e=e_1$ to restore the correct weight $n_1$ for the segment joining two 
such insertion points when this segment winds around a cylinder. 
In general, electro-magnetic operators with electric charge $e$ and magnetic charge 
$m$ have conformal dimension
\eqn\confdi{ h_{e,m}={e(e-2e_1)\over 4g}+{g\over 4}m^2 }
provided $(e-e_1) m=0$, where $g$ is the coupling of the free field, 
with $n_1=-2\cos (\pi g)$, i.e. $g=1-e_1$. 
In our case, the operator of insertion of $1$ line originating from one
endpoint corresponds to having $m=1/2$ and $e=e_1$ (see for instance \DaDu).
Taking now $n_1=0$, i.e. $e_1=1/2$, we identify for $\psi_{\pm 1}$ the conformal 
dimension
\eqn\confdimop{ h_1=h_{e_1,1/2}= -{e_1^2\over 4 (1-e_1)} +{1-e_1\over 16}=-{3\over 32}}
and its gravitationally dressed counterpart through \twokpz:
\eqn\dressedone{ \Delta_1=
{{1\over 2}\sqrt{11}-\sqrt{5} \over \sqrt{29}-\sqrt{5}} \ .}
The semi-meander generating function \semimean\ then reads\foot{When $n_1,n_2\to 0$,
we have to pick the term proportional to $n_1n_2$ in $\langle \phi_1\phi_1\rangle$, namely 
compute $\lim_{n_1,n_2\to 0}\langle \phi_1\phi_1\rangle/(n_1n_2)$. By a slight abuse
of notation, we still write the result as $\langle \phi_1\phi_1\rangle$.}
\eqn\smgen{ \langle \phi_1 \phi_1\rangle \sim (x_c-x)^{2\Delta_1-\gamma_{str}} }
as a particular case of \correla\
and gives the semi-meander asymptotics
\eqn\asyme{ {\bar M}_n \sim {R^{n} \over n^{\bar \alpha}}\ ,} 
where
\eqn\aldet{ {\bar \alpha}= 1 -\gamma_{str}(-4) +2 \Delta_1=
1+{1\over 24}\sqrt{11}(\sqrt{29}+\sqrt{5})=2.05319873... }
This value again is in agreement with the recent numerical estimate 
${\bar \alpha}=2.056(10)$
of \Goli.

\newsec{More river geometries}

The above picture leads to many interesting results for $n_1=n_2=0$
using operators of insertion
of more lines. In the matrix model formalism, we introduce
\eqn\highop{ \phi_k =\lim_{N\to \infty} {1\over N}\sum_{i=1}^{n_1}{\rm Tr}(B_i^k) }
creating $k$ black lines from a point.
The magnetic analogue of this operator has dimension
\eqn\dimcon{ h_k=h_{e_1=1/2,k/2}={k^2-4 \over 32} }
and we get the dressed dimension 
\eqn\drek{ \Delta_k={{1\over 2}\sqrt{8+3 k^2}- \sqrt{5} \over \sqrt{29}-\sqrt{5}}\ . }

\fig{Meandric numbers with particular river configurations: ``k-star"=the
river is a star with $k$ branches, with $k$ univalent vertices and one $k$-valent one;
``p,k-star"=the river is made of two stars with respectively $p$ and $k$ branches
one of which is common to both, i.e. with one $p$-valent, 
one $k$-valent and $p+k-2$ univalent
vertices; ``pentagon"=the river is a tree with three trivalent vertices and five
univalent ones;
``cherry"=the river is made of a loop connected to a segment, with one tri-valent
and one univalent vertices; ``eight"=the river is made of two loops connected
at one point. On the sphere, we may send the central vertex of the river
of the ``$k$-star" graphs to infinity, yielding $k$ parallel semi-infinite rivers
connected at infinity. 
Moreover, the ``cherry" configuration is equivalent to that
of one loop including one segment, and the connection point may be sent to
infinity, yielding a semi-infinite river parallel to an infinite one.
Similarly, the ``eight" configuration
can be transformed into that of two included loops, and 
the connection point may be again sent to
infinity, yielding two parallel rivers (with all their ends
connected at infinity).}{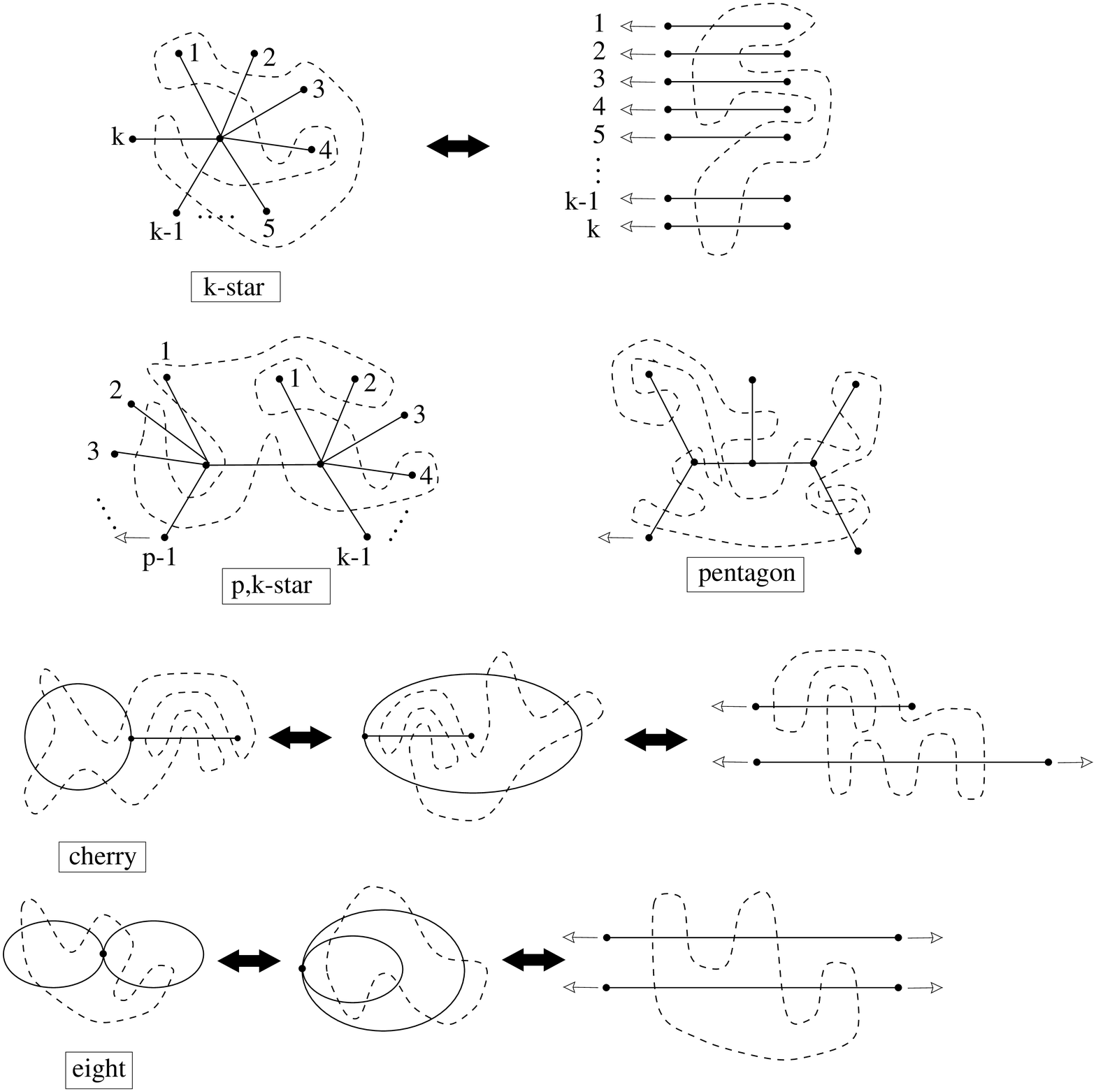}{10.5cm}
\figlabel\star

We may use the operators $\phi_k$ to compute a class of generalized meandric
numbers, corresponding to situations when the river is made of several connected 
segments and/or loops, for instance in the cases of the river configurations 
depicted in Fig.\star: 
\eqn\reso{\eqalign{
``k-star":&\langle (\phi_1)^k \phi_k\rangle 
\sim (x_c-x)^{\alpha_{k-star}-1}\cr
&\Rightarrow \ \ \alpha_{k-star}=k\Delta_1+\Delta_k-\gamma_{str}+2-k \cr
``p,k-star":&\langle (\phi_1)^{p+k-2} \phi_p \phi_k\rangle 
\sim (x_c-x)^{\alpha_{p,k-star}-1}\cr
&\Rightarrow \ \ \alpha_{p,k-star}=(p+k-2)\Delta_1+\Delta_p+\Delta_k-\gamma_{str}+3-p-k \cr
``pentagon":&\langle (\phi_3)^3(\phi_1)^5 \rangle
\sim (x_c-x)^{\alpha_{pentagon}-1}\cr
&\Rightarrow \ \ \alpha_{pentagon}=5 \Delta_1+3 \Delta_3-\gamma_{str}-5\cr
``cherry":&\langle \phi_1 \phi_3 \rangle\sim (x_c-x)^{\alpha_{cherry}-1}\cr
&\Rightarrow \ \ \alpha_{cherry}=\Delta_1+\Delta_3-\gamma_{str}+1 \cr
``eight":&\langle \phi_4 \rangle \sim (x_c-x)^{\alpha_{eight}-1}\cr
&\Rightarrow \ \ \alpha_{eight}=\Delta_4-\gamma_{str}+2 \cr}}
where again $\langle ...\rangle$ stands for $\lim_{n_1,n_2\to 0} 
\langle ...\rangle/(n_1n_2)$.
In all these cases, the corresponding meandric numbers count the configurations
of a single road crossing the connected river graph, one vertex of
which  we have sent to infinity. As illustrated in Fig.\star, we have chosen to
send the central $k$-valent vertex of the $k$-star to infinity say on the left,
leaving us with  a river formed of $k$ parallel half-lines (connected at infinity);
in that case, the counting function for theses configurations is $k\langle \phi_k
\phi_1^k\rangle$, to account for the $k$-fold degeneracy. 
Note also that the ``cherry" river configuration equivalently corresponds to
a segment included in a loop, and that the connecting point can
be sent to infinity, leaving us with a configuration of a semi-infinite
river parallel to an infinite one, as shown in Fig.\star. 
Similarly, the ``eight" configuration corresponds to two parallel infinite rivers. 

In particular, eqn.\reso\ yields for the ``3-star", ``pentagon",
``cherry" and ``eight" configurations
of river, corresponding respectively to $k=3$ in the first case of \reso, and the
three last ones:  
\eqn\starwars{\eqalign{ 
\alpha_{3-star}&={1\over 2}{{3}\sqrt{11} +\sqrt{35}-8\sqrt{5}\over \sqrt{29}-\sqrt{5}}
+{5+\sqrt{145}\over 12} -1 \cr
&=-1+{1\over 48}(\sqrt{5}+\sqrt{29})(3\sqrt{11}+\sqrt{35}-4\sqrt{5})=0.09899483... \cr
\alpha_{pentagon}&={1\over 2}{{5}\sqrt{11} +3\sqrt{35}-16\sqrt{5}\over \sqrt{29}-\sqrt{5}}
+{5+\sqrt{145}\over 12} -5\cr
&=-5+{1\over 48}(\sqrt{5}+\sqrt{29})(5\sqrt{11}+3\sqrt{35}-12\sqrt{5})=-3.80941298... \cr
\alpha_{cherry}&={1\over 2}{\sqrt{11}+\sqrt{35}-4\sqrt{5}\over \sqrt{29}-\sqrt{5}}
+{5+\sqrt{145}\over 12} +1\cr
&=1+{1\over 48}(\sqrt{5}+\sqrt{29})(\sqrt{11}+\sqrt{35}) =2.46592898... \cr
\alpha_{eight}&={\sqrt{14}-\sqrt{5}\over \sqrt{29}-\sqrt{5}} 
+{5+\sqrt{145}\over 12} +2 \cr
&=2+{1\over 24}(\sqrt{5}+\sqrt{29})(\sqrt{14}+\sqrt{5})=3.89823486... \cr}}

In addition to the cases of Fig.\star, we may also allow for disconnected rivers.
For instance, we may realize a river made of $k$ distinct segments
by considering, for $n_1,n_2\to 0$: 
\eqn\consid{\eqalign{
\lim_{n_1,n_2\to 0} &{1\over n_1^kn_2} \langle (\phi_1)^{2k}\rangle\sim
(x_c-x)^{\alpha_{k-segments}-1} \cr
&\Rightarrow \ \ \alpha_{k-segments}=2k\Delta_1-\gamma_{str}+3-2k \ ,\cr}}
whereas we may get a river made of $k$ distinct segments, plus a $p$-star
by considering
\eqn\dicons{\eqalign{
\lim_{n_1,n_2\to 0} &{1\over n_1^{k+1}n_2} \langle (\phi_1)^{2k+p}\phi_p\rangle\sim
(x_c-x)^{\alpha_{k-segments+p-star}-1} \cr
&\Rightarrow \ \ \alpha_{k-segments+p-star}=
(2k+p)\Delta_1+\Delta_p-\gamma_{str}+3-2k-p \ .\cr}}
In both cases \consid\-\dicons, and contrary to the situations of Fig.\star\ where
the river is connected (possibly through the point at infinity), the
road {\it must} visit all the connected components of river, to ensure that the
meandric black and white graph is globally connected.

Note finally that for $k=2$ in \drek\ we get $\Delta_2=0$ (this is no surprise,
as $h_2=0$ and $\phi_2$ is the dressed identity operator, also called puncture
operator in gravity). This means that the insertion of any number of $\phi_2$
in a correlation has the same effect as the same number of applications
of $x{d\over dx}$ on the critical asymptotics. More precisely, we have
\eqn\puncture{
{\langle X (\phi_2)^p \rangle \over \langle X \rangle } \sim
(x_c-x)^{-p} } 
for any combination of operators $X$.
The puncture operator $\phi_2$ can therefore
be viewed as that of marking a point on the river. In particular, we have
\eqn\meanalter{ \langle \phi_2\rangle \propto \sum_{n\geq 1} M_n x^{2n} \sim
(x_c-x)^{1-\gamma_{str}}}  
in agreement with $\alpha=2-\gamma_{str}$.

\newsec{Discussion}

The value \charcen\ of the central charge $c(n_1,n_2)$ allows us to extend the 
meander results to arbitrary values of $n_1,n_2$. Defining the multi-river
and multi-road meander polynomial
\eqn\mepol{ m_{n}(n_1,n_2)=4 n f_{2n}(n_1,n_2) \ , }  
we have the following prediction for its large $n$ asymptotics:
\eqn\predic{\eqalign{
m_{n}(n_1,n_2) &\sim {R(n_1,n_2)^{2n}\over n^{\alpha(n_1,n_2)}}\cr 
\alpha(n_1,n_2)&= 2-{1\over 12}\bigg(c(n_1,n_2)-1-\sqrt{(25-c(n_1,n_2))
(1-c(n_1,n_2))}\bigg)\ .\cr}}
This can be checked against the exact result \CK\ in the case $n_2=1$ and $n_1$
arbitrary, where $e_2=1/3$, obtained from the solution of the
$O(n_1,n_2=1)$ matrix model in the limit of large size $N$. The result of \CK\ reads 
\eqn\rres{ R(n_1,1)= 2 {\sin^2(\pi {e_1\over 2})\over e_1^2},  
\qquad \alpha(n_1,1)=2+{e_1\over 1-e_1} \ ,}
and this value agrees with our general prediction \predic.

Note that we have no definite answer for $R(n_1,n_2)$, as the critical value $x_c$
of the cosmological constant is a non-universal quantity, expected to depend
on $n_1$ and $n_2$ explicitly, and not just on $c(n_1,n_2)$.
We expect \predic\ to hold only if
\eqn\valid{ c(n_1)\leq 1, \qquad c(n_2)\leq 1, \quad {\rm and}\quad c(n_1,n_2)\leq 1}
Indeed, the $O(n)$ model is no longer critical for $n>2$ ($c(n)\leq 1$ for $n\leq 2$), 
therefore
we expect a different phase whenever $n_1>2$ or $n_2>2$. This phase
has been investigated in the case of the gravitational $O(n)$ model
\EKR\ and found to have $\gamma_{str}=+1/2$
uniformly. Moreover, the relation
\kpz\ breaks down when $c>1$, and the gravity is then known \BAR\
to degenerate in such a
way that surfaces with long fingers dominate (branched polymer phase), and 
throughout this phase one has $\gamma_{str}=+1/2$. The corresponding value of
the meander exponent is therefore $\alpha=2-\gamma_{str}=3/2$.
This is in
agreement with the results for the exact large $n_2$ expansion
of the meander polynomial $n_2 m_n(n_1=0,n_2)$ denoted by $m_n(n_2)$ in \NOUS.

The operator of insertion of one black line is still well defined and has
conformal dimension $h_1$ given in \confdimop, and dressed counterpart
\eqn\drecou{ \Delta_1(n_1,n_2)=
{\sqrt{{3\over 2}(1-e_1)-c(n_2)}-\sqrt{1-c(n_1,n_2)}\over
\sqrt{25-c(n_1,n_2)}-\sqrt{1-c(n_1,n_2)}} \ . }
This leads to the asymptotics of the multi-river and multi-road semi-meander
polynomial
${\bar m}_n(n_1,n_2)= {\bar f}_n(n_1,n_2) $
in which the river is made of one segment and a number of loops. 
Recalling that the numbers ${\bar m}_n(n_1,n_2)$ are generated by the correlation
function $\langle \phi_1 \phi_1 \rangle$ of the $O(n_1,n_2)$ model,
we find
\eqn\asysemi{\eqalign{
{\bar m}_n(n_1,n_2)&\sim{R(n_1,n_2)^n \over n^{{\bar \alpha}(n_1,n_2)}}\cr 
{\bar \alpha}(n_1,n_2)&= \alpha(n_1,n_2)-1+2 \Delta_1(n_1,n_2) \cr
&=1+{1\over 24}\big(\sqrt{25-c(n_1,n_2)}+\sqrt{1-c(n_1,n_2)}
\big) \sqrt{6(1-e_1)-4c(n_2)}\ . \cr    }}

In the particular case $n_2=1$ (with $c(n_2)=0$), we have
\eqn\partism{ {\bar \alpha}(n_1,1)=1+{1\over 24}\bigg((2-e_1)\sqrt{6\over 1-e_1}
+e_1\sqrt{6\over 1-e_1}\bigg) \sqrt{6(1-e_1)}={3\over 2} }
independent of the value of $n_1$. This agrees with the known result for $n_1=0$,
where ${\bar m}_n(0,1)=c_n=(2n)!/((n+1)!n!)\sim 4^n/n^{3/2}$, in terms
of the Catalan numbers $c_n$. 

We expect the formula \asysemi\ to hold only if
\eqn\hold{ c(n_1,n_2)\leq 1, \qquad c(n_1)\leq 1, \quad {\rm and}\quad
c(n_2)\leq {3\over 2}(1-e_1) }
The last bound corresponds to ${\bar \alpha}(n_1,n_2)=1$ in 
\asysemi, and actually corresponds to a ``winding transition" beyond
which semi-meander configurations with large numbers of circles (with only one
intersection with the river) dominate.
\medskip
\noindent{\bf Acknowledgements:} We thank F. David for useful discussions.

\medskip

\noindent{\bf Note added in proof:$\ \ $ }
Since we implicitly use {\it magnetic} operators in our derivation
of the exponents $\alpha$ for the various geometries, the edges of the
corresponding graphs are implicitly oriented. These orientations must
be such that all the edges meeting at a given vertex have the
same inwards or outwards orientation. All the graphs we considered here
can be equipped with such orientations except for the geometry of
the ``eight" and that of the ``cherry". For these two cases, the apparent
problem can be bypassed by {\it marking} a point on {\it each} closed loop,
thus adding a bivalent vertex across which the orientation is reversed.
Strictly speaking, the values of $\alpha_{eight}$ and $\alpha_{cherry}$ of
Eq. (5.5) thus correspond to $\alpha_{eight}=\alpha_{eight}^{marked}+2$
and $\alpha_{cherry}=\alpha_{cherry}^{marked}+1$ where the superscript
indicates that configurations are counted with a marked point on each loop.
It is easy to prove that, without these marked points, the exponent
for the ``eight" (resp. ``cherry") geometry simply reduces to the meander (resp.
semi-meander) exponent, hence  $\alpha_{eight}^{unmarked}=\alpha$
given by Eq. (1.2) and $\alpha_{cherry}^{unmarked}=\bar \alpha$ given by
Eq. (1.3).

\listrefs \bye